# Lead-free room-temperature ferroelectric thermal conductivity switch using anisotropies in thermal conductivities

Lucile Féger[1], Carlos Escorihuela-Sayalero[2], Jean-Michel Rampnoux[3], Kyriaki Kontou[4], Micka Bah[1], Jorge Íñiguez-González[5,6], Claudio Cazorla[2], Isabelle Monot-Laffez[1], Sarah Douri[4,7], Stéphane Grauby[3], Riccardo Rurali[8] ****, Stefan Dilhaire[3] ***, Séverine Gomès[4] **, Guillaume F. Nataf[1] *

[1] GREMAN UMR7347, CNRS, University of Tours, INSA Centre Val de Loire, 37000 Tours, France
[2] Departament de Física, Universitat Politècnica de Catalunya, Campus Nord B4-B5, Barcelona 08034, Spain
[3] Université de Bordeaux, CNRS, LOMA, UMR 5798, F-33400 Talence, France
[4] Univ Lyon, CNRS, INSA-Lyon, Université Claude Bernard Lyon 1, CETHIL UMR5008, F-69621, Villeurbanne, France
[5] Materials Research and Technology Department, Luxembourg Institute of Science and Technology (LIST), L-4362 Esch-sur-Alzette, Luxembourg
[6] Department of Physics and Materials Science, University of Luxembourg, L-4422 Belvaux, Luxembourg
[7] Laboratoire National de Métrologie et d'Essais (LNE), 29, Avenue Roger Hennequin, 78197 Trappes, France
[8] Institut de Ciència de Materials de Barcelona, ICMAB-CSIC, Campus UAB, 08193 Bellaterra, Spain
* guillaume.nataf@univ-tours.fr
** severine.gomes@insa-lyon.fr
*** stefan.dilhaire@u-bordeaux.fr
**** rrurali@icmab.es

**Materials with on-demand control of thermal conductivity are the prerequisites to build thermal conductivity switches, where the thermal conductivity can be turned ON and OFF. However, the ideal switch, while required to develop novel approaches to solid-state refrigeration, energy harvesting, and even phononic circuits, is still missing. It should consist of an active material only, be environment friendly, and operate near room temperature with a reversible, fast, and large switching ratio. Here, we first predict by *ab initio* electronic structure calculations that ferroelectric domains in barium titanate exhibit anisotropic thermal conductivities. We confirm this prediction by combining frequency-domain thermoreflectance and scanning thermal microscopy measurements on a single crystal of barium titanate. We then use this gained knowledge to propose a lead-free thermal conductivity switch without inactive material, operating reversibly with an electric field. At room temperature, we find a switching ratio of $1.6 \pm 0.3$, exceeding the performances of state-of-the-art materials suggested for thermal conductivity switches.**

The ability to control electron flow with logic units such as switches, diodes, and transistors, has been a life changer in our society. However, achieving a similar control of heat flows is still a challenge, due to the intrinsic nature of heat-carrying phonons that are extremely difficult to manipulate with external fields. Thus, an active management of heat flow is still missing, even though it would enable efficient systems for energy recovery, conversion, and transport [1–6]. Many theoretical and experimental works have been dedicated to tuning thermal transport in solid materials, within the framework of phononics [7–12]. They rely mostly on mass-transport [13,14], phase transitions [14–17], and topological defects [18–22], driven by different stimuli: temperature [15], light [16], electric potential [14], magnetic field [18], strain field [19], electric field [15,17,20–22]. Nevertheless, investigated materials fail to exhibit both a large thermal conductivity switching ratio and a fast response time.

Among these materials, ferroelectrics appear as promising for future developments due to the fast response (approximately nanoseconds) of their domain structures to electric fields [23,24]. So far, most thermal conductivity switches, i.e., elements where the thermal conductivity can be tuned reversibly by an external applied field, rely on the interaction between phonons and interfaces between ferroelectric domains (called domain walls). Samples with high densities of domain walls exhibit a reduced thermal conductivity ($\kappa_{low}$) compared to samples with low densities of domain walls ($\kappa_{high}$). At low temperature (<10 K), large switching ratios ($R=\kappa_{high}/\kappa_{low}$ up to 5) are obtained [25,26] but these cannot be upheld at ambient conditions where the phonon mean free path decreases below 100 nm [26,27], which is about one to two orders of magnitude smaller than the typical distance between adjacent domain walls in bulk materials. A solution consists in working with thin films in which the distance between domain walls is smaller, but it comes at the cost of adding a thick substrate whose thermal conductivity is fixed. Furthermore, while different samples can exhibit reasonably different thermal conductivities in thin films, reaching ratios of up to 2.8 [19,28], the difficulties encountered in drastically changing the domain structure with an electric field in practice lead to much smaller effective thermal conductivity ratios (<1.3) [15,20–22].

Here, we develop an alternative approach based on the structural differences between ferroelectric domains, which – in contrast with phonon scattering at domain walls – is weakly sensitive to the decrease in phonon mean free path upon increasing temperature. We present first-principles calculations predicting different thermal conductivity values depending on the direction of heat propagation with respect to the ferroelectric polar axis. We then combine local thermal conductivity measurements by frequency-domain thermoreflectance (FDTR) and scanning thermal microscopy (SThM) on ferroelectric domains in a single crystal of barium titanate (BaTiO$_3$) to quantify the difference in thermal conductivity. Finally, we show that when an electric field is applied, the ferroelectric domains reorient, yielding a thermal conductivity ratio of 1.6 at room temperature, thus creating a thermal conductivity switch that exceeds the performance of state-of-the-art alternatives.

**Simulations of anisotropic thermal conductivity**. At room temperature, BaTiO$_3$ is ferroelectric and has a tetragonal structure, with space group *P4mm*. After structural optimization from first principles, including volume expansion at 300 K within the quasiharmonic approximation, we obtain lattice parameters $a = b = 3.969$ Å (short axis) and $c = 4.065$ Å (long axis), and a polarization $P = 30.8$ µC cm$^{-2}$, in good agreement with experiments [29]. To calculate the thermal conductivity of BaTiO$_3$ at room temperature (300 K), we compute the finite-temperature renormalized phonon dispersion at 300 K and the third-order force constants, and then solve the Boltzmann transport equation (BTE) beyond the relaxation time approximation (RTA). We obtain different room temperature values depending on the direction of the heat propagation: $\kappa_a = \kappa_b = 4.5$ W m$^{-1}$ K$^{-1}$ and $\kappa_c = 3.7$ W m$^{-1}$ K$^{-1}$ [with the short (*a* and *b*) and long (*c*) axes labeled as above; Fig. 1a), corresponding to a thermal conductivity ratio of 1.2. In line with previous calculations on similar ferroelectric perovskites (e.g., PbTiO$_3$, Refs. [30,31]), we thus find a lower value of the thermal conductivity along the polarization direction. We performed similar calculations on a broader temperature range, from 280 to 350 K, and find that the anisotropy of the thermal conductivity is maintained (Fig. 1a).

To elucidate the origin of this anisotropy, we analyze the cumulative contribution to the thermal conductivity as a function of phonon frequency (Fig. 1b), which shows that the difference in thermal conductivity between $\kappa_a = \kappa_b$ and $\kappa_c$ is dominated by phonons with $f > 4$ THz and $f < 10$ THz.

Next, we analyze separately the harmonic and anharmonic properties — respectively, phonon velocities and phonon lifetimes — that make up the thermal conductivity (see Eq. 1 in Methods). Figure 1c displays the *a* and *c* components of the velocity of each phonon. Here it can be seen that, while no significant difference is appreciated for low-frequency and high-frequency phonons, the phonons in the mid-frequency range present larger *a* than *c* component of the velocity. Next, we look at the influence of anharmonic processes on the anisotropy. To this end, we weight the computed phonon lifetimes as $\tau_i = \tau \cdot q_i/|\mathbf{q}|$, where $i = a, b, c$ and $\mathbf{q}$ is the phonon wavevector, so that, e.g., $\tau_a$ is the effective lifetime of phonons that propagate along the *a* axis. In Fig. 1d we plot the difference $\Delta\tau = \tau_a - \tau_c$, which turns out to be mostly positive, meaning that phonons propagating along *a* (or *b*) live often longer than the *c* propagating ones. This difference is particularly sizable at low frequencies. Therefore, both phonon velocities and phonon lifetimes concur in making the components of the thermal conductivity within the plane orthogonal to the polarization, $\kappa_a = \kappa_b$, larger than the one parallel to it, $\kappa_c$.

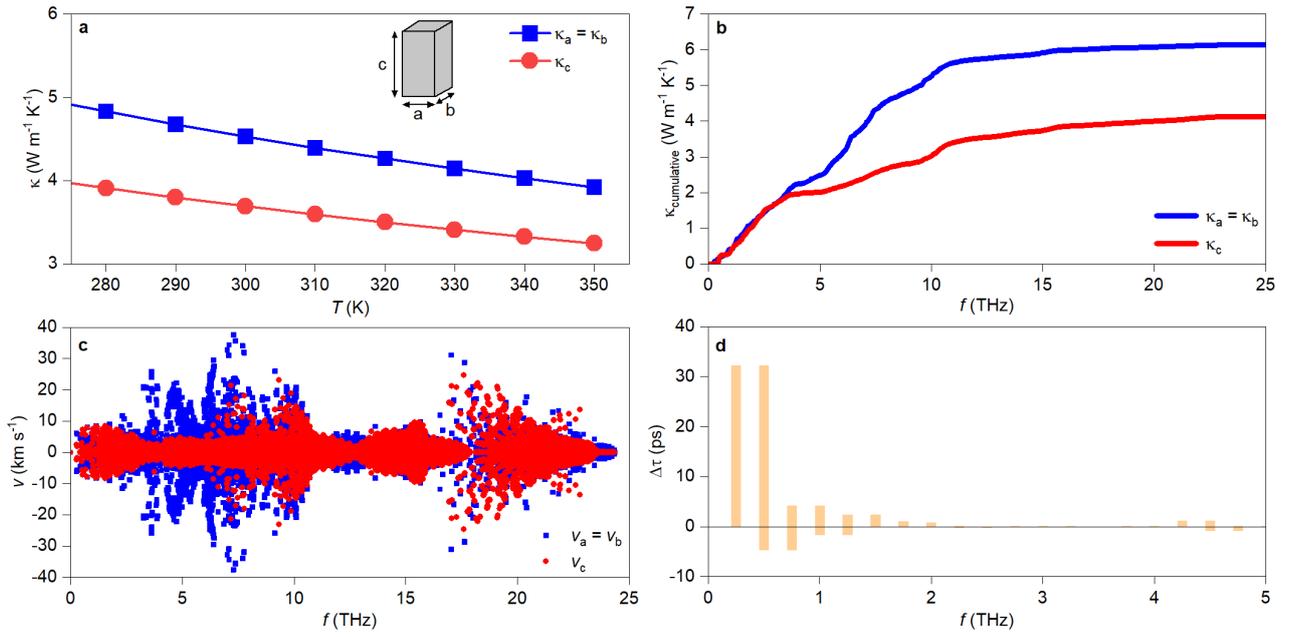

**Figure 1. Thermal anisotropy.** (a) Thermal conductivity $\kappa$ as a function of temperature. The inset schematizes the BaTiO$_3$ tetragonal unit cell. (b) Cumulative thermal conductivity as a function of frequency. (c) Phonon velocities as a function of frequency $f$. (d) Difference between the lifetimes $\tau_a = \tau_b$ and $\tau_c$, averaged over frequency intervals of 0.25 THz, where $\tau_i$ is defined as $\tau \cdot q_i/|\mathbf{q}|$. The analysis in panels (b) and (d) is carried out at the RTA level of the theory.

**Ferroelectric domain structure.** To demonstrate experimentally the dependence of the thermal conductivity on the direction of the heat propagation with respect to the polar axis, we perform local thermal conductivity measurements on ferroelectric domains in a (001)-oriented single crystal of BaTiO$_3$ (SurfaceNet GmbH). Figure 2a shows an optical image in reflection of the sample, with the cantilever of the piezoresponse force microscope visible. Domains appear as stripes parallel to the (010) direction, which indicates by symmetry a succession of domains where the ferroelectric polarization alternates between in-plane and out-of-plane orientations. This is confirmed by the topographic signal from the piezoresponse force microscopy (PFM) measurement performed on the same area, where factory rooflike features typical of a succession of in-plane and out-of-plane domains are observed (Fig. 2b) [32]. The orientation of the polarization is then deduced from the PFM in-plane and out-of-plane phase signals depicted in Fig. 2c and d. In Fig. 2c, obtained by vertical PFM, domains where the phase signal is noisy indicate in-plane domains, called *a* domains, while a clear 180° phase-contrast is observed in out-of-plane domains, called *c* domains. In Fig. 2d, obtained by lateral PFM, *a* domains appear with a clear 180° phase-contrast, while *c* domains appear noisy. Thus, from left to right, upward slopes from figure 2b correspond to domains where the ferroelectric polarization lies in-plane (Fig. 2d), while downward slopes indicate out-of-plane orientations (Fig. 2c). The sense of the polarization in the out-of-plane domains has been obtained by comparing the PFM signal with the signal obtained on a BiFeO$_3$ film of known polarization sense [33] and is shown in Fig. 2c. This organization of domains is the same throughout the sample, even though some areas also exhibit herringbone patterns.

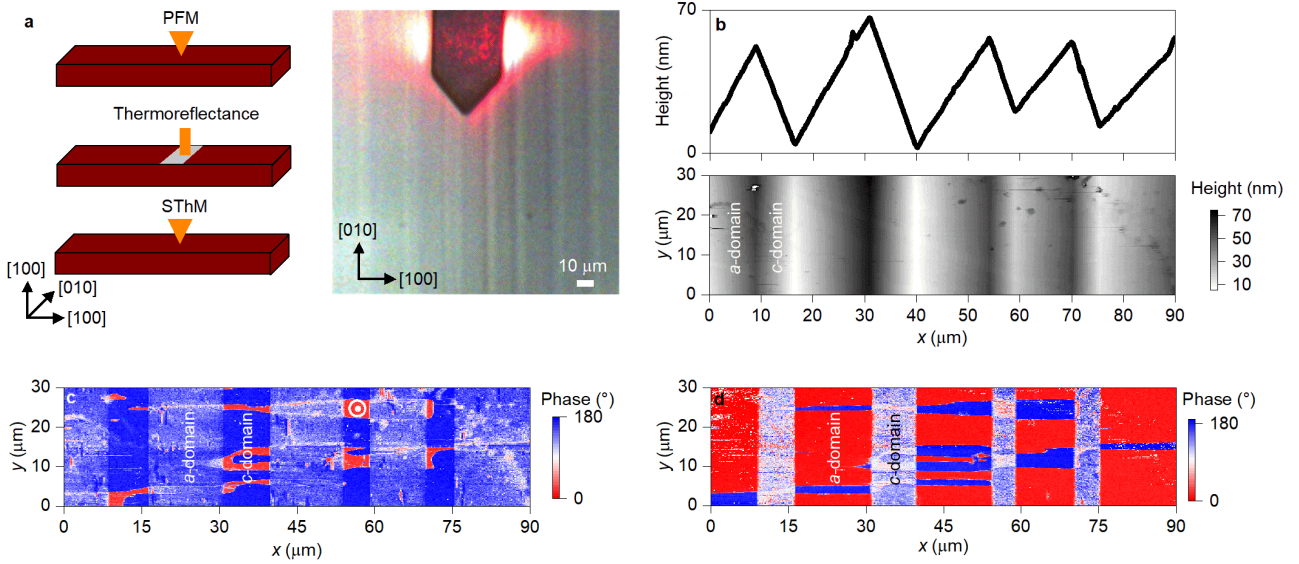

**Figure 2. Ferroelectric domain structure. a** Optical image revealing ferroelectric-ferroelastic domains in a (001)-oriented BaTiO$_3$ single crystal. The schematics illustrate the measurement geometry for PFM, FDTR and SThM, where the orange triangles indicate the tips, the orange rectangle the laser beams, and the grey rectangle the aluminium film. **b** Topography as obtained by atomic force microscopy. **c** Phase of the vertical piezoresponse force microscopy signal revealing out-of-plane domains. The circles indicate the sense of the ferroelectric polarization. **d** Phase of the lateral piezoresponse force microscopy signal revealing in-plane domains.

**Spatially resolved thermal conductivity measurements.** To evaluate the thermal conductivity of the ferroelectric domains in the single crystal, local FDTR measurements are performed at room temperature on domains covered with a thin aluminium film and whose orientations were previously identified by PFM (schematics in Fig. 2a). Six measurements using beam diameters about 12 µm are performed on in-plane and out-of-plane domains. The optimization process is implemented in the frequency range between 5 and 100 MHz for which the thermal conductivity ratio is not dependent on the other thermal and geometrical parameters (see Supplemental Material SI-I [34]). Indeed, the model is responsive to the aluminium thickness and the volumetric heat capacity of the material. However, these data are known and serve as inputs to the model. A parametric analysis of the optimization results is performed to verify that the thermal conductivity ratio is almost constant according to probable values of these model input parameters. The experimental measurements confirm that *a* domains have a higher cross-plane thermal conductivity, $\kappa_{a,b} = 3.9 \pm 0.5$ W m$^{-1}$ K$^{-1}$, than *c* domains $\kappa_c = 2.6 \pm 0.3$ W m$^{-1}$ K$^{-1}$. The anisotropy degree is $\kappa_{a,b}/\kappa_c = 1.5 \pm 0.3$. The uncertainty is obtained with a propagation of the experimental noise using the input parameter uncertainties associated with the aluminium film thickness, the beam diameters, the beam offset, and the volumetric heat capacities. A detailed sensitivity analysis is presented in the Supplemental Material (SI-I.2).

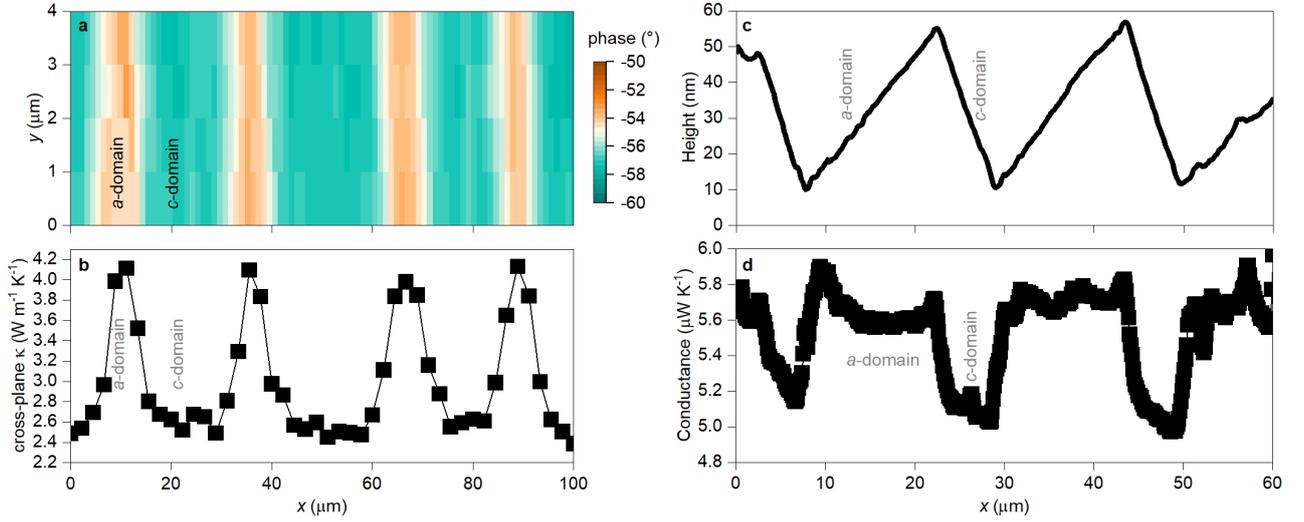

**Figure 3. Local thermal conductivity measurements. a** FDTR map of the phase signal at 5 MHz showing the periodic change correlated to the ferroelectric domain type change (*a* domain = in-plane, *c* domain = out-of-plane). **b** Cross-plane thermal conductivity identified on 9 different domains by FDTR following a linear scan of 100 µm. **c** Topography profiled measured simultaneously as the thermal signal of the scanning thermal microscope. **d** Profile of the thermal conductance calculated from the output voltage.

To confirm the reproducibility of the measurements, a 100 x 4 µm$^2$ mapping is carried out (Fig. 3a). The phase response at 5 MHz, which is directly related to the material thermal conductivity, shows the correlation between the direction of the polarization in the ferroelectric domains and the change in the thermal conductivity. The profile of the cross-plane thermal conductivity is extracted from a set of phase images at different frequencies in the [5 MHz, 100 MHz] range (Fig. 3b), providing consistent results with previous measurements.

Offering volumetric spatial resolution of only a few cubic micrometers at maximum for bulk samples investigated in ambient air conditions, ambient-air SThM is then used to confirm these results. Two signals are simultaneously obtained as a function of the probe location on the sample surface: the topography and the voltage variation, $\Delta V_{\text{out}}$, which depends on the thermal conductivity of the locally probed volume. From $\Delta V_{\text{out}}$, a profile of the variation of the probe mean temperature is deduced and converted into a thermal conductance by using a calibration curve (see SI-II for details about the SThM probe calibration). Figure 3c shows the topography and Fig. 3d the thermal conductance profiles obtained from SThM on the area previously imaged by PFM. There is a clear correlation between the domains identified by the topography and the thermal conductance, with sharp changes when moving from one domain to the other. It reveals that in-plane and out-of-plane ferroelectric domains exhibit different thermal conductances: $5.6 \pm 0.1$ µW K$^{-1}$ for in-plane domains (*a* domains) and $5.0 \pm 0.1$ µW K$^{-1}$ for out-of-plane domains (*c* domains), as shown in Fig. 3d.

A three-dimensional model of the thermal transport in the BaTiO$_3$ single crystal was then developed using Finite elements modeling to calculate the absolute values of in-plane and cross-plane thermal conductivities in the ferroelectric domains (see SI-II for details about the thermal conductivity component estimation). In agreement with our first-principles simulations and the FDTR measurements, the thermal conductivity of each domain is described by an anisotropic thermal conductivity with components $k_a = k_b$ and $k_c$ (in the *a*, *b*, and *c* directions respectively). With our measurement geometry (Fig. 1a, 2a), in *c* domains, $\kappa_c$ is the cross-plane thermal conductivity, while $\kappa_{a,b}$ is the in-plane thermal conductivity. In *a* domains, it is the opposite: $\kappa_{a,b}$ is the cross-plane thermal conductivity while $\kappa_c$ is the in-plane thermal conductivity. In SThM, $\kappa_{a,b}$ and $\kappa_c$ are identified for both *c* domains and *a* domains simultaneously. We obtain $k_c = 2.7 \pm 0.4$ W m$^{-1}$ K$^{-1}$ and $k_{a,b} = 4.2 \pm 0.3$ W m$^{-1}$ K$^{-1}$. These conductivities are consistent with average thermal conductivities reported in the literature for BaTiO$_3$ (ref. [35]) and in excellent agreement with our FDTR measurements. The anisotropy degree is found to be of $\kappa_{a,b}/\kappa_c = 1.6 \pm 0.3$, in agreement with our FDTR measurements. It is however higher than the ratio of 1.2 found with *ab initio* calculations, even though the calculated values of the thermal conductivity are in good agreement. Yet, computing the anisotropy degree is a much more delicate task, especially with so small thermal conductivities. Indeed, even a slight variation in the absolute value of one of the components, e.g., within the range of variability associated to the choice of the exchange-correlation functional [36], can significantly affect the anisotropy degree. Additionally, in our first-principles calculations only anharmonicity up to the third order is considered and we cannot rule out that higher-order scattering processes are comparatively more anisotropic. Also, we consider ideal, defect-free single-crystal, which is normally a very stringent standard, even for samples of the highest purity.

**Discussion.** Our results indicate that in BaTiO$_3$ the thermal conductivity depends on the direction of propagation of heat with respect to the electric polarization direction. In addition, it is known that the direction of the electric polarization can be changed by applying an electric field that leads to the rearrangement of ferroelectric domains with a fast response (approximately nanoseconds) [23,24]. We thus propose to build a thermal conductivity switch based on the intrinsic thermal anisotropy of the ferroelectric domains, instead of relying on domain walls or phase transitions as currently found in the literature. We depict, in Fig. 4, a schematic of such thermal conductivity switch. Our optical images in transmission show a succession of *a* and *c* domains. Under the application of an electric field, the *c* domain vanishes and comes back at almost the same position when the electric field is removed. Below, we show the corresponding variations in thermal conductivity in the cross-plane direction, as deduced from our local thermal conductivity measurements: *c* domains exhibit a low cross-plane thermal conductivity while *a* domains exhibit a high cross-plane thermal conductivity. It illustrates the mechanism of a local thermal conductivity switch controlled by an electric field, with a thermal conductivity ratio corresponding to the anisotropy degree of $1.6 \pm 0.3$ measured previously.

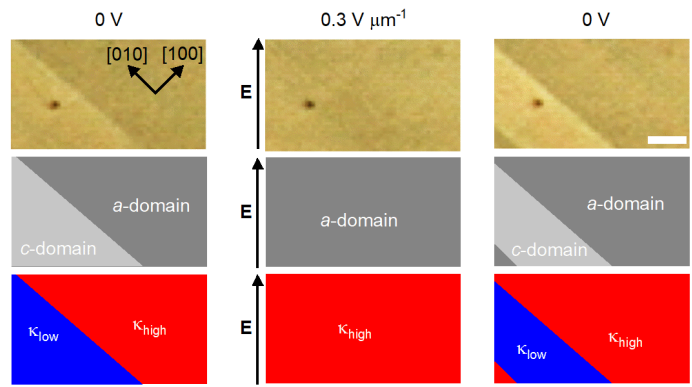

**Figure 4. A thermal conductivity switch.** Optical images in transmission of a (001)-oriented single crystal of barium titanate under the application of an electric field. Schematics of the ferroelectric domain structure observed and the corresponding thermal conductivity in the cross-plane direction, demonstrating the principle of a thermal conductivity switch. The scale bar corresponds to 10 μm.

In Fig. 5, the thermal conductivity ratio is compared to the performance of state-of-the-art room temperature solid-state thermal conductivity switches based on the application of an electric field. Most of them rely on the interaction between phonons and domain walls, with ratios close to 1.3. There are probably two reasons why higher ratios could not be reached. First, obtaining a low thermal conductivity state requires a very dense pattern of domain walls, such that the distance between consecutive domain walls is close to the mean free path of the phonons at room temperature. In ferroelectrics, this is usually achieved only in thin films [37] on specific substrates, and even in these systems, domains sizes stay around 100 nm [19,28]. Second, to obtain a high thermal conductivity state the domain structure must drastically change under the application of an electric field (ideally to reach a state free of domain walls), which is in fact difficult in thin films. In bulk materials, ratios close to 1.3 have only been obtained by applying electric field over long periods of time (several minutes).

The main advantage of our proposed thermal conductivity switch is that it is based on the structural difference between ferroelectric domains and, thus, it does not require a high density of domain walls. Furthermore, it relies on environment friendly materials (avoiding the presence of lead as in most cited examples) and it is a bulk effect, thus maximizing the amount of active material compared to thin films.

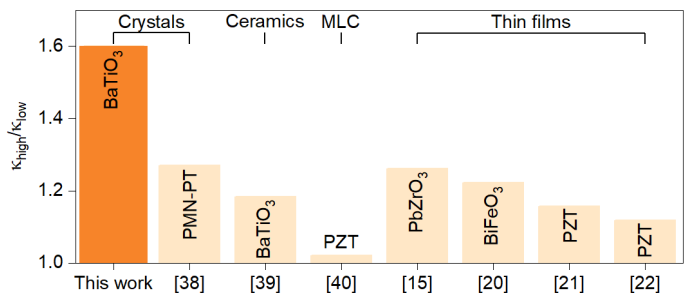

**Figure 5.** Thermal conductivity ratio for our work (orange) and for state-of-the-art solid-state thermal conductivity switches using an electric field at room temperature. MLC = multilayer capacitor. PZT = Pb(Zr,Ti)O$_3$.

As a final comment, let us note that in all previous thermal conductivity switches based on ferroelectric materials, the domain structure is changing [15,19–22,28,38–40] and our results demonstrate that such change strongly alters the thermal conductivity. If half of the ferroelectric polarization of a sample moves from in-plane to out-of-plane, the spatially averaged thermal conductivity is divided by a ratio of ~1.3, similar to the values reported in the literature, but often attributed to changes in densities of domain walls. It may thus be useful to revisit the interpretation of previous work, taking into account the possible role of anisotropy in the conductivity of individual domains.

In summary, we combined *ab initio* electronic structure calculations and local thermal conductivity measurements to reveal a large difference in thermal conductivity depending on the direction of propagation of heat with respect to the ferroelectric polarization in a single crystal of BaTiO$_3$. This difference can be used to build an efficient thermal conductivity switch based on the intrinsic thermal-transport anisotropy of ferroelectric domains. The same effect should be observed in other ferroelectrics, and may even be enhanced in perovskites with larger ferroelectric distortions, paving the way for a new generation of thermal conductivity switches at room temperature.

## Methods

### First-principles calculations

We performed density functional theory calculations with the VASP code [41] and projector augmented waves [42] with an energy cutoff of 520 eV and the generalized-gradient approximation PBEsol [43]. The following electrons were explicitly treated as valence: Ba $6s^25p^65s^2$, Ti $3d^44s^23p^6$, and O $2s^2 2p^4$. The Brillouin zone was sampled with a 12x12x12 **k**-points grid and the atomic positions were optimized until all the atomic forces were smaller than 0.005 eV·Å$^{-1}$. The examined tetragonal phase of BaTiO$_3$ was described by a five atoms unit cell. The corresponding electric polarization was calculated with the Born effective charges method [44].

The second-order interatomic force constants were obtained with the DynaPhoPy code [45]. In particular, the anharmonic lattice dynamics of tetragonal BaTiO$_3$ (i.e., *T*-renormalized phonons) was determined from *ab initio* molecular dynamics (AIMD) simulations [46]. This method essentially consists in applying a normal-mode decomposition technique in which the atomic velocities generated during the AIMD simulation runs are projected into a basis of phonon eigenvectors. The involved AIMD simulations were carried out in the (*N*, *V*, *T*) ensemble. The simulated temperature was kept fluctuating around a set-point value by using a Nosé-Hoover thermostat. We used simulation boxes containing *N*=135 atoms and periodic boundary conditions were applied along the three Cartesian directions. Newton's equations of motion were integrated by using the customary Verlet's algorithm with a time-step length of 1.5 fs. Γ-point sampling for integration within the first Brillouin zone was employed and the total duration of the AIMD simulations was of 115 ps.

We used the Phonopy code [47] to calculate the thermal expansion of tetragonal BaTiO$_3$. To this end, we considered five different volume points near the equilibrium one obtained at 0 K conditions; then we computed the *T*-renormalized phonons and anharmonic lattice dynamics at each of those volume points. By doing so, we could determine the equilibrium volume of the tetragonal BaTiO$_3$ to be 65.7 Å$^3$ at *T*=300 K.

We computed the third-order IFCs by finite differences in a 3×3×3 supercell by means of the thirdorder.py code [48] with a cutoff up to sixth neighbors in the three-phonon scattering processes, which yields converged values of the thermal conductivity. The harmonic and anharmonic IFCs obtained were used as inputs to solve the phonon BTE beyond the RTA on a 26×26×26 grid of **q** points with the almaBTE code [49]. Scattering from isotopic disorder is accounted for through the model of Tamura [50]. The lattice thermal conductivity is computed as

$$\kappa_{ij} = \frac{1}{k_B T \Omega N} \sum_\lambda f_\lambda (f_\lambda + 1)(h\nu_\lambda)^2 \tau_\lambda v_{i,\lambda} F_{j,\lambda} \quad (1)$$

where $k_B$, $T$, $\Omega$, and $N$ are the Boltzmann's constant, Planck's constant, temperature, volume of the unit cell, and number of **q** points, respectively. The sum runs over all phonon modes, the index λ including both **q** point and phonon band. $f_\lambda$ is the equilibrium Bose-Einstein distribution function, $\tau_\lambda$ is the lifetime, $\nu_\lambda$ and $v_{i,\lambda}$ are the frequency and the group velocity of phonon λ. $F_{j,\lambda}$ is a mean free displacement that, at the RTA level, corresponds to the mean free path.

These calculations assume a perfect, defect-free, bulklike single crystal and neglect anharmonic effects beyond third-order or electron-phonon scattering, though we do not expect the latter to play a significant role due to the insulating character of BaTiO$_3$.

**Piezoresponse force microscopy**

PFM is a useful tool to image at nanoscale the orientation of the electrical polarization in ferroelectric materials [51]. PFM is based on the inverse piezoelectric effect where the sample contracts/expands when subjected to an electric field. The tip-sample interaction informs about the piezoelectric coefficient ($d_{33}$ in the vertical direction and $d_{15}$ in the horizontal one) and the direction/sense of the electrical polarisation. PFM studies are performed using a Bruker Dimension Icon atomic force microscope (AFM). Prior to measurements, the sample is sonicated in acetone and ethanol before being glued with silver paste on a metallic support. The PFM setup enables the surface topography and polarization orientation to be simultaneously recorded by scanning a biased conductive tip in contact mode across the sample surface. While the tip is grounded, an ac bias of 3000 mV at a drive frequency of 15 kHz is applied to the sample. The scan rate is set to 0.1 Hz while a constant and low force of 72 nN is applied to the tip. A diamond coated silicon tip (DDESP-V2, purchased from Bruker) is used due to its toughness and its sensitive nanoelectrical performance.

**Thermoreflectance**

FDTR [52–54] uses the temperature dependence of the reflectance of a material to optically measure the thermal properties of it. This experiment is frequency oriented with a pump laser beam that generates the material excitation. The laser intensity is modulated, usually with a sinus shape. To probe the material response, another laser beam named probe is reflected at its surface. A set of measurements with a frequency sweep permits to reconstruct the thermal impulse response of the material. Finally, a performed optimization between a thermal model and the measurements provides thermal characteristics of the material.

Firstly, a thin metal film is deposited on top of the sample. Its goal is to do an in-depth confinement of the absorbed heat and to enable the surface temperature sensing by its reflectance change. Here, aluminium electrodes are sputtered on the BaTiO$_3$ surface by the physical vapor deposition process. The relevant parameters for the deposition are the electrical power (300 W), the argon flow rate (22 SCCM; SCCM denotes cubic centimeter per minute at STP), the chamber vacuum (5×10$^{-3}$ Torr) and the deposition time (90s). The thickness of the deposited Al electrode is checked from the height measurement using an AFM and is found to be ~ 110 nm. Secondly, the sample is put under a microscope objective to focus and to make the positioning of the pump and probe beams. The intensity of the continuous wave pump laser is electrically modulated with a frequency sweep. The absorbed energy induces a periodic temperature change. Thus, the continuous wave laser probe overlaps the heating area and is reflected to a photodiode. The relative variation of the probe intensity is detected with a lock-in synchronized on the pump frequency. In our case, under a 10X microscope objective, the Gaussian pump and probe $1/e^2$ radii are 13 and 11 µm respectively. This indicates the lateral spatial resolution. The pump and probe wavelengths are 808 and 786 nm respectively. For a given position on the sample, to determine its thermal response, the pump frequency is swept from a few kilohertz up to a few hundred megahertz.

To evaluate the thermal conductivity, an optimization is performed between the FDTR phase measurements and a quadrupole model. The implemented model [52,55] is axisymmetric as the laser beam profile and each layer of the sample is represented by a transfer matrix. It considers the thicknesses of layers, the thermal conductivities, the volumetric heat capacities, and the thermal resistance between the layers. Moreover, the optical penetration of the pump beam is modelled as well as the Gaussian beam profile of the pump and probe.

**Scanning thermal microscopy**

In its conventional mode, SThM uses a small, self-heated resistive tip, which is put in contact with the surface of the investigated sample and allows probing the material thermal properties in the subsurface volume of the sample just below the probe-sample contact [56]. Depending on the size of the tip, the probed volume can vary but in air it remains lower than few µm$^3$ when performing experiments with a tip with a curvature radius lower than 100 nm [56]. The SThM setup used in this work is based on an atomic force microscopy instrument (NTEGRA AURA - NT-MDT). The SThM probe used is the KNT probe (from Kelvin NanoTechnology). It is a silicon nitride AFM probe where the tip is a triangular platform on which is deposited a thin resistive palladium (Pd) ribbon. Gold pads are sputtered on the cantilever to make possible the electrical measurement of the probe response by means of a thermal control unit.

For the measurements of thermophysical properties, the Pd ribbon, which is the probe sensor, is heated by Joule effect and acts as a sensor thermometer/heat fluxmeter and a local heating source for the sample simultaneously. In this work, homemade electronics based on an unbalanced Wheatstone bridge is used for the measurement of the electrothermal response of the probe. Thermal raw data are the voltage variation $\Delta V_{\text{out}}$ associated with the Wheatstone bridge imbalance. From the $\Delta V_{\text{out}}$ value one can obtain the probe voltage $V_p$, the electrical current heating the sensor $I_p$ and the electrical resistance of the probe $R_p$. In this work, the electrical current in the probe is of the order of 1 mA. A calibration in different steps is however needed for thermal measurement in controlled conditions [57]. Measurements are performed in ambient air conditions.

**Optical microscopy**

We visualized the ferroelectric domain structure with an optical microscope in transmission mode (Olympus BX60), using an objective with magnification 20X and numerical aperture 0.4, and a polarizer and an analyzer in crossed configuration. The optical contrast arises from the unique orientation of the optical indicatrix for each ferroelectric domain state. In order to move domain walls, we applied an electric field along the $[110]_{pc}$ direction. We cycled the voltage (electric field) between 0 and $\pm$ 0.3 V $\mu$m$^{-1}$ at 40 V s$^{-1}$ following a standard ferroelectric hysteresis loop.

**Data availability**

All data needed to evaluate the conclusions in the paper are available in this paper or its Supplemental Material. The data that support the findings of this study are available from the corresponding authors upon reasonable request.

**Acknowledgements**

We thank Flavien Barcella, Raphaël Doineau and Béatrice Negulescu for technical support for aluminium sputtering. We thank Juan Carlos Acosta Abanto for technical support for developing 3D simulation for SThM measurement post-treatment. This paper was cofunded by the European Union [European Research Council (ERC), DYNAMHEAT, No. 101077402; European Program FP7-NMP-LARGE-7, QUANTIHEAT, No. 604668]. Views and opinions expressed are, however, those of the authors only and do not necessarily reflect those of the European Union or the ERC. We acknowledge financial support by MCIN/AEI/10.13039/501100011033 under Grant No. PID2020-119777GB-I00, the "Ramón y Cajal" fellowship RYC2018-024947-I and the Severo Ochoa Centres of Excellence Program (CEX2019-000917-S), by the Generalitat de Catalunya (2021 SGR 01519) and by the Luxembourg National Research Fund (C21/MS/15799044/FERRODYNAMICS). Some of this work has been performed on the CERTeM (microelectronics technological research and development center) technological platform supported by French region Centre Val de Loire, Tours Metropolis and FEDER funds.

**Author contributions**

GFN, SG and SD conceived the study. GFN and LF prepared the samples for the different measurements and performed optical microscopy measurements. CES, JÍ, CC, and RR performed *ab* *initio* calculations. MB performed PFM measurements. JMR, StG, and SD performed thermoreflectance measurements. LF and KK performed SThM experiments. SG and SaD made the post-treatment of SThM experimental data. All authors contributed to the discussion of the results. All authors contributed to the preparation of the manuscript.